\documentclass[aps,twocolumn,groupedaddress,superscriptaddress,nofootinbib,floatfix]{revtex4-1}
\usepackage{graphicx}
\usepackage{color}
\usepackage{pifont}
\usepackage{amsmath,amsfonts,amssymb}

\newcommand{\met}{\mbox{$\protect \raisebox{.3ex}{$\not$}E_T$}}

\DeclareMathOperator{\arctanh}{arctanh}

\begin{document}

\vspace*{-3cm}
\vspace*{0.5cm}
\begin{flushright}
IFT-UAM/CSIC-21-123 \\
\end{flushright}
\vspace{0.cm}

\title{A template method to measure the $t \bar t$ polarisation}

\author{J. A. Aguilar-Saavedra}
\affiliation{Instituto de F\'\i sica Te\'orica, IFT-UAM/CSIC, c/ Nicolás Cabrera 13-15, 28049 Madrid}
\affiliation{Departamento de F\'{\i}sica Te\'{o}rica y del Cosmos, Universidad de Granada, 18071 Granada, Spain}

\author{M.C.N. Fiolhais}
\affiliation{Science Department, Borough of Manhattan Community College, The City University of New York,  
     199 Chambers St, New York, NY 10007, USA}
\affiliation{The Graduate School and University Center, The City University of New York, 365 Fifth Avenue, New York, NY 10016, USA}     
\affiliation{LIP, Physics Department, University of Coimbra, 3004-516 Coimbra, Portugal}

\author{P. Mart\'in-Ramiro}
\affiliation{Instituto de F\'\i sica Te\'orica, IFT-UAM/CSIC, c/ Nicolás Cabrera 13-15, 28049 Madrid}

\author{J. M. Moreno}
\affiliation{Instituto de F\'\i sica Te\'orica, IFT-UAM/CSIC, c/ Nicolás Cabrera 13-15, 28049 Madrid}

\author{A. Onofre}
\affiliation{LIP, Physics Department, University of Coimbra, 3004-516 Coimbra, Portugal}
\affiliation{Centro de F\'{\i}sica das Universidades do Minho e do Porto (CF-UM-UP), Universidade do Minho, 4710-057 Braga, Portugal}

\begin{abstract}
We develop a template method for the measurement of the polarisation of $t \bar t$ pairs produced in hadron collisions. The method would allow to extract the individual fractions of $t_L \bar t_L$, $t_R \bar t_R$, $t_L \bar t_R$ and $t_R \bar t_L$ pairs with a fit to data, where $L,R$ refer to the polarisation along any axis. These polarisation fractions have not been independently measured at present. Secondarily, the method also provides the net polarisation of $t$ and $\bar t$, as well as their spin correlation for arbitrary axes.
\end{abstract}

\maketitle

\section{Introduction}

The measurement of the top quark properties started shortly after its discovery at the Tevatron~\cite{CDF:1995wbb,D0:1995jca}. The high statistics achieved at the Large Hadron Collider (LHC) has provided us with a huge dataset of single (anti-)top and $t \bar t$ pairs, which can be exploited for precision measurements in the search for any departure from the predictions of the Standard Model (SM). And this will be even more the case at the high-luminosity upgrade (HL-LHC). With such large statistics, the main source of uncertainties in the comparison between theory and experiment are the experimental systematic uncertainties, as well as theoretical uncertainties due to higher-order corrections in perturbation theory~\cite{Czakon:2020qbd}. The latter are currently being reduced by two-loop calculations; the former may be reduced, not only with a better knowledge of the detectors, but also with alternative methods to perform the measurements. 

The purpose of this paper is to demonstrate the feasibility of a template method to measure the polarisation of the $t \bar t$ pairs produced at the LHC, namely the relative fractions $a_{LL}$, $a_{RR}$, $a_{LR}$, $a_{RL}$, corresponding to the production of $t_L \bar t_L$, $t_R \bar t_R$, $t_L \bar t_R$, and $t_R \bar t_L$,  where the $R,L$ subscripts refer to the polarisation along a given axis (not necessarily the helicity). For this feasibility study we use the dilepton decay channel of the $t \bar t$ pair, $t \bar t \to \ell^+ \nu b \, \ell^- \nu \bar b$, where the background is relatively small, especially if one restricts the selection to events where the charged leptons have different flavor and requires a $b$-tagged jet. Notice that, at present, only the $t$ and $\bar t$ polarisations, as well as the $t \bar t$ spin correlation, which are linear combinations of the $a_{XX'}$ (with $X,X' = L,R$), have been measured. The template method here proposed allows for an independent determination of these quantities, which could also allow to test for P- and CP-violating effects.

As it will be detailed in Section~\ref{sec:2}, the template method relies on generating samples corresponding to the four above polarisation combinations and extracting the coefficients $a_{XX'}$ from a fit to the measured distribution. Once the effect of hadronisation, detector resolution, kinematical reconstruction of the $t$ and $\bar t$ momenta, and phase space cuts are suitably incorporated (details are discussed in Section~\ref{sec:3}), the {\it parton-level} coefficients $a_{XX'}$ can be extracted by a fit of the measured sample to a combination of the simulated templates. Detailed results are presented in Sections~\ref{sec:4} and \ref{sec:5}, and Section~\ref{sec:6} is devoted to a brief discussion of our results.

\section{The template method}
\label{sec:2}

The template method is based on the expansion of the $t \bar t$ cross section, which can be written as
\begin{equation}
\sigma = \sigma_{RR} +  \sigma_{LL} + \sigma_{RL} + \sigma_{LR} \,,
\label{ec:sexp}
\end{equation}
where $R,L$ refer to the polarisations of the top quark and anti-quark along some direction, which does not need to be the same for both particles.
Although quantum interference effects do exist between polarisation states, this expansion can be promoted to the differential level as long as one does not consider distributions sensitive to that interference~\cite{Aguilar-Saavedra:2012bvs,Aguilar-Saavedra:2014yea}. In particular, to disentangle the different polarisation contributions one can consider, for each decaying quark, the angle $\theta_\ell$ between the charged lepton $\ell = e,\mu$ in the top rest frame and the top spin direction, which follows the well-known distribution
\begin{equation}
\frac{1}{\sigma} \frac{d\sigma}{d\cos\theta_\ell} = \frac{1}{2}(1 + \kappa_\ell \cos \theta_\ell) \,,
\end{equation}
with $\kappa_{\ell^+} = - \kappa_{\ell^-} = 1$ in the SM at leading order (LO). (Next-to-leading order (NLO) corrections are at the permille level~\cite{Brandenburg:2002xr}.)
Let us define the short-hand notation $z_1 \equiv \cos \theta_{\ell^+}$, $z_2 \equiv \cos \theta_{\ell^-}$. 
Integrating out the rest of kinematical variables, and defining for convenience
\begin{align}
& f(z_1,z_2) = \frac{1}{\sigma} \frac{d\sigma}{dz_1 dz_2} \,, \notag \\
& f_{XX'}(z_1,z_2) = \frac{1}{\sigma_{XX'}} \frac{d\sigma_{XX'}}{dz_1 dz_2} \,,
\label{ec:f}
\end{align}
the normalised doubly-differential distribution can be expanded, at the parton level, as
\begin{eqnarray}
f(z_1,z_2) = \sum_{XX'} a_{XX'} f_{XX'}(z_1,z_2) \,, 
\label{ec:expansion}
\end{eqnarray}
with $a_{XX'} = \sigma_{XX'} / \sigma$ the coefficients we want to determine, which obey the normalisation condition 
\begin{equation}
\sum_{XX'} a_{XX'} = 1
\end{equation}
from their definition. We remark that it is precisely the integration over lepton azimuthal angles $\phi_{\ell^+}$, $\phi_{\ell^-}$ in the top (anti-)quark rest frame that cancels interference terms, ensuring the validity of the expansion (\ref{ec:expansion}).
Phase-space cuts differently affect the various polarisation contributions, and one can write
\begin{equation}
\frac{d\bar\sigma}{dz_1 dz_2} = \sum_{X,X'} \frac{d\bar\sigma_{XX'}}{dz_1 dz_2} + \dots \,,
\end{equation}
with the bar denoting the quantities after cuts and the dots standing for interference terms, which do not identically cancel due to phase space cuts. Note however that with mild cuts on transverse momenta and pseudo-rapidities, the interference effects are still unimportant. 
Defining the overall efficiencies $\varepsilon = \bar \sigma / \sigma$, $\varepsilon_{XX'} = \bar \sigma_{XX'} / \sigma_{XX'}$ and  $\bar f(z_1,z_2)$, $\bar f_{XX'}(z_1,z_2)$ in analogy to (\ref{ec:f}), it follows that
\begin{equation}
\varepsilon \bar f(z_1,z_2) = \sum_{XX'} a_{XX'} \varepsilon_{XX'} \bar f_{XX'}(z_1,z_2) + \Delta_\text{int}(z_1,z_2) \,,
\label{ec:expansion3}
\end{equation}
with a (small) interference term $\Delta_\text{int}$.
Detector resolution and reconstruction effects make this term relevant for an accurate determination of $a_{XX'}$. The reason is that, if the angles $\theta_{\ell^+}$ and $\theta_{\ell^-}$ are not well determined, the mismatch in the reconstruction of these angles further prevents the cancellation of the interference terms by integration over $\phi_{\ell^+}$ and $\phi_{\ell^-}$. The effect can be taken into account by including a residual interference term $\Delta_\text{int}$ in the expansion, which in first approximation can be evaluated in the SM using Monte Carlo simulation.\footnote{Should data be incompatible with the SM, the term $\Delta_\text{int}$ could also be calculated in an iterative process. However, as we will find in Section~\ref{sec:5}, the SM evaluation of $\Delta_\text{int}$ works well even to extract the polarisation parameters in a non-SM sample where we introduce a top chromomagnetic dipole moment that yields $t \bar t$ spin correlations incompatible with current measurements.}
Summarising, the effects of the detector simulation and reconstruction are encoded in
\begin{itemize}
\item[(i)] Different efficiencies for the different polarisation components, as well as for the inclusive sample. In an actual measurement, they must be calculated with Monte Carlo simulation.
\item[(ii)] Different functions $\bar f(z_1,z_2)$ and $\bar f_{XX'}(z_1,z_2)$. The latter are the templates generated with Monte Carlo simulation, while the former is measured in data to extract the polarisation components $a_{XX'}$.
\item[(iii)] A correction $\Delta_\text{int}$ that corrects the linear combination of templates for residual interference effects, and is determined from Monte Carlo, in principle assuming the SM.
\end{itemize}
To conclude this section, let us mention that the determination of the $a_{XX'}$ with a fit to the binned $\bar f(z_1,z_2)$ distribution is more precise if the event selection cuts on transverse momenta ($p_T$) of the charged leptons and $b$ quarks are not very stringent --- otherwise the differences between the templates are partly washed out. We have performed a simple test by considering the templates for different helicity combinations without any hadronisation and detector simulation, but with kinematical cuts on $p_T$. The difference between two functions $f(z_1,z_2)$, $g(z_1,z_2)$ can be measured by the quantity
\begin{equation}
\rho(f,g) = \frac{||f-g||^2}{||f|| \, ||g||} \,,
\end{equation}
with the usual norm $||f||^2 = \int |f(z_1,z_2)|^2 \, dz_1 dz_2$.\footnote{Note that the templates defined as in (\ref{ec:f}) are normalised in the sense $\int f(z_1,z_2) dz_1 dz_2 = 1$.} The distances between templates, evaluated using a $20\times 20$ grid for the binning of the distributions, are collected in Table~\ref{tab:dist} for several choices of parton-level cuts. The effect of a lower cut on the $b$ quark $p_T$ is minimal, while the cut on lepton $p_T$ erases some of the difference between polarisation combinations..

\begin{table}[htb]
\begin{center}
\begin{tabular}{cccccc}
& no cut & $p_T^\ell \geq 30$ & $p_T^{\ell,b} \geq 30$ & $p_T^{\ell(b)} \geq 50(30)$
\\[1mm]
$\rho(f_{LL},f_{RR})$ & 1.5  & 1.29 & 1.29 & 1.15
\\[1mm]
$\rho(f_{LL},f_{LR})$ & 1    & 0.80 & 0.79 & 0.73
\\[1mm]
$\rho(f_{LL},f_{RL})$ & 1    & 0.85 & 0.83 & 0.71
\\[1mm]
$\rho(f_{RR},f_{LR})$ & 1    & 0.80 & 0.79 & 0.73
\\[1mm]
$\rho(f_{RR},f_{RL})$ & 1    & 0.85 & 0.83 & 0.71
\\[1mm]
$\rho(f_{LR},f_{RL})$ & 1.5  & 1.34 & 1.31 & 1.23
\end{tabular}
\caption{Normalised distance between the templates for different helicity combinations at the parton level, without kinematical cuts and with cuts on the transverse momenta of the charged lepton and $b$ quark (transverse momenta are in GeV).}
\label{tab:dist}
\end{center}

\end{table}

\section{$t \bar t$ simulation and reconstruction}
\label{sec:3}

The event samples with definite polarisation are generated using {\scshape Protos}~\cite{protos}, using the {\scshape HELAS}~\cite{Murayama:1992gi} implementation of polarisation projectors discussed in Ref.~\cite{Aguilar-Saavedra:2014yea}. As a basis for the spins we use the one introduced in Ref.~\cite{Bernreuther:2015yna}. For the top quark:
\begin{itemize}
\item K-axis (helicity): $\hat k_t$ is a normalised (unit modulus) vector in the direction of the top quark three-momentum in the $t \bar t$ rest frame.
\item R-axis: $\hat r_t$ is in the production plane and defined as $\hat r_t = \mathrm{sign}(y_p) (\hat p_p - y_p \hat k_t)/r_p$, with $\hat p_p = (0,0,1)$ the direction of one proton in the laboratory frame, $y_p = \hat k_t \cdot \hat p_p$, $r_p = (1-y_p^2)^{1/2}$. Note that the $\hat r_t$ is the same if we use for the definition the direction of the other proton $- \hat p_p$.
\item N-axis: $\hat n_t$ is orthogonal to the production plane and defined as $\hat n_t = \mathrm{sign}(y_p) (\hat p_p \times \hat k_t)/r_p$, which again is independent of the proton choice.
\end{itemize}
For the anti-quark the axes are $\hat k_{\bar t} = - \hat k_t$, $\hat r_{\bar t} = - \hat r_t$, $\hat n_{\bar t} = - \hat n_t$.

For the sake of brevity, in this paper we restrict ourselves to templates with the same spin axis for the top quark and antiquark (that is, for both we consider either K, R or N). Spin correlations with mixed directions can be studied and have been experimentally measured~\cite{ATLAS:2016bac,CMS:2019nrx,Aaboud:2019hwz}. For each set of axes (KK, RR, NN) for the top quark and anti-quark, we generate with {\scshape Protos} four event samples
\begin{equation*}
t_R \bar t_R \,,\quad t_L \bar t_L \,,\quad  t_R \bar t_L \,,\quad t_L \bar t_R \,, 
\end{equation*}
where $R$ and $L$ refer to the spin along the chosen axis. Each sample has $1.5 \times 10^6$ events, totalling 18 million events. The LO NNPDF 2.1~\cite{Ball:2011mu} parton distribution functions (PFDs) are used, with fixed factorisation scale $Q=m_t$. 

A SM $t \bar t$ sample with $1.5 \times 10^6$ events, as well as a non-SM sample of the same size but with a large top chromomagnetic dipole moment (CMDM) $d_V = 0.036$, are generated with {\scshape  MadGraph} at the LO. In this case we use NNPDF 2.3~\cite{Ball:2013hta} PDFs and dynamic factorisation scale $Q = (m_t^2 + p_T^2)^{1/2}/2$. The CMDM is introduced as an interaction 
\begin{equation}
\mathcal{L} = - \frac{g_s}{m_t} \bar t \sigma^{\mu \nu} (d_V + i d_A \gamma_5) \frac{\lambda^a}{2} t G_{a}^{\mu \nu} \,,
\end{equation}
in standard notation with $g_s$ the strong coupling constant, $\lambda^a$ the Gell-Mann matrices and $G_a^{\mu \nu}$ the gluon field strength tensor. A possible chromoelectric dipole moment $d_A$ is set to zero. The Lagrangian is implemented in {\scshape Feynrules}~\cite{Alloul:2013bka} and interfaced to {\scshape MadGraph5} using the universal Feynrules output~\cite{Degrande:2011ua}.

Hadronisation and parton showering is performed with {\scshape Pythia~8}~\cite{Sjostrand:2007gs} and detector simulation with {\scshape Delphes 3.4}~\cite{deFavereau:2013fsa} using the configuration for the CMS detector. The reconstruction of jets and the analysis of their substructure is done using {\scshape FastJet}~\cite{Cacciari:2011ma}. Jets are reconstructed using the anti-$k_T$ algorithm~\cite{Cacciari:2008gp} with a radius $R=0.4$.

Events are required to have two isolated charged leptons (electrons or muons) with pseudo-rapidity $|\eta| \leq 2.4$, the leading one with $p_T \geq 25$ GeV and the sub-leading one $p_T \geq 20$ GeV. Jets are required to have $|\eta| < 2.4$ and $p_T \geq 30$ GeV. We apply two different reconstruction methods, which we summarise in turn.

\subsection{Reconstruction method I}
\label{sec:recI}

\begin{figure*}[htb]
\begin{center}
\begin{tabular}{ccc}
\includegraphics[width=5.8cm,clip=]{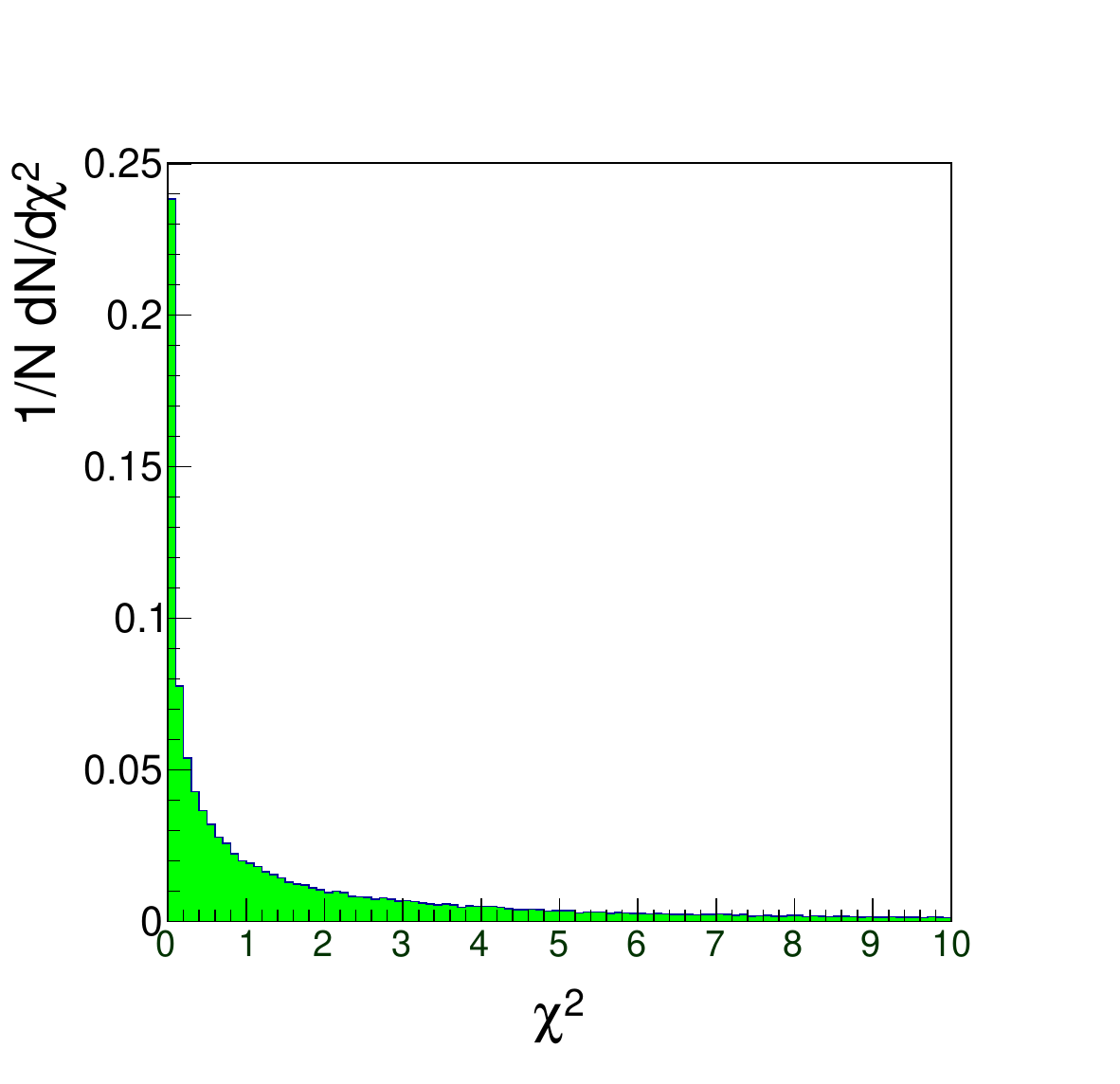} &
\includegraphics[width=5.8cm,clip=]{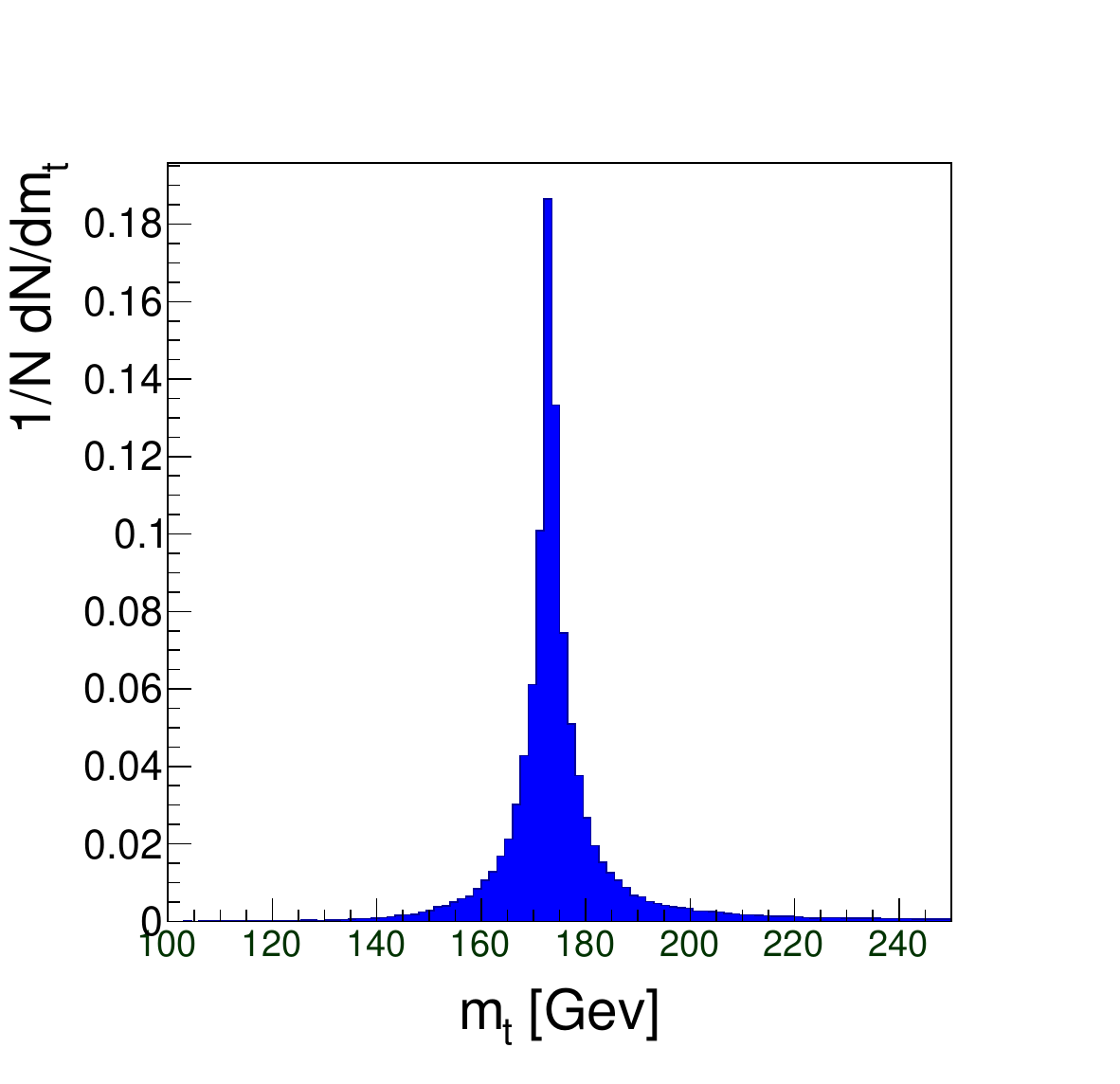} &
\includegraphics[width=5.8cm,clip=]{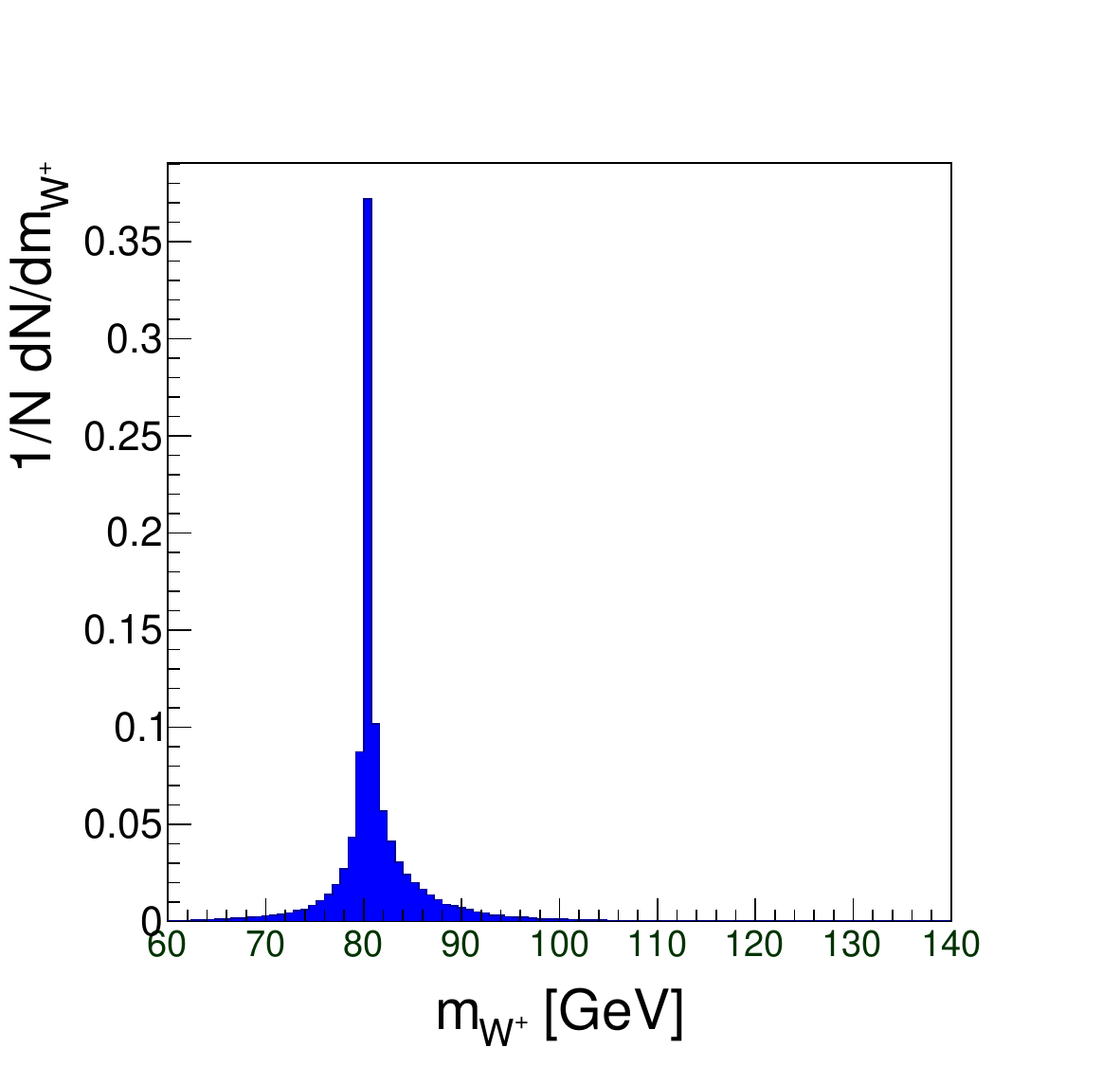} 
\end{tabular}
\caption{The $\chi^2$ (left), top quark mass (centre) and $W$ boson mass (right) normalised distributions are represented, after reconstruction.}
\label{fig:rec}
\end{center}
\end{figure*}

In this method, a scan is performed over the four-momenta of the undetected neutrinos, assuming that the missing transverse energy $\met$ is originated by these undetected particles, that is,
\begin{align}
& (\met)_x = (p_{\nu_1})_x + (p_{\nu_2})_x \,, \notag \\
& (\met)_y = (p_{\nu_1})_y + (p_{\nu_2})_y \,.
\end{align}
In total, there are four unknowns to scan over, for example, $(p_{\nu_1})_x$, $(p_{\nu_1})_y$, $(p_{\nu_1})_z$ and $(p_{\nu_2})_z$. The momenta of the top quarks and $W$ bosons are reconstructed as
\begin{eqnarray}
& p_{W^+} = p_{l^+} + p_{\nu_1} \,, \nonumber \\
& p_{W^-} = p_{l^-} + p_{\nu_2} \,, \nonumber \\
& p_{t} = p_{l^+} + p_{\nu_1} + p_{b_1} \,, \nonumber \\
& p_{\bar t} = p_{l^-} + p_{\nu_2} + p_{b_2} \,,
\label{ec:prec}
\end{eqnarray}
with the different four-momenta correspond to the leptons, neutrinos and b-quarks in the final state.
The transverse momentum of the top quark ($p_{t}^{T}$) is expected to compensate the anti-top quark transverse momentum ($p_{\bar t}^{T}$). A global $\chi^2$ is defined for each event according to
\begin{eqnarray}
\chi^2 & = & \frac{(m_t^{\textrm{rec}}-m_t)^2}{\Gamma_t^2} +             
         \frac{(m_{\bar{t}}^{\textrm{rec}}-m_{\bar{t}})^2}{\Gamma_{\bar{t}}^2} +
         \frac{(m_{W+}^{\textrm{rec}}-m_W)^2}{\Gamma_{W+}^2} \notag \\
         & & + \frac{(m_{W-}^{\textrm{rec}}-m_W)^2}{\Gamma_{W-}^2} +             
         \frac{(p_t^T - p_{\bar{t}}^T)^2}{(\sigma_p^T)^2}\, ,
\label{eq:chi2}
\end{eqnarray}
where $m_{W} = 80.4$ GeV, $m_{t} = 173$ GeV, $\Gamma_{t,\bar t} = 11.5$ GeV, $\Gamma_{W^\pm} = 7.5$ GeV and $\sigma_p^T = 20$ GeV, represent typical experimental resolutions for the reconstructed width of the top quarks and $W$ bosons, and top quark transverse momentum, respectively. Ultimately, these numbers can also be viewed as weighting factors of the global $\chi^2$, giving different importance to its individual terms. In order to find the best solution for each event i.e. the minimum value for the $\chi^2$, performing a loop over all the available jets and leptons to find the best combination among that minimises the global $\chi^2$, as is usually done at the LHC. The normalised distributions for the $\chi^2$, the top quark mass and the $W^+$ boson mass, are represented in Figure~\ref{fig:rec} after reconstruction and for the best solution found. For the top anti-quark and the $W^-$ boson the distributions are similar.

A loose selection is applied to events, requiring $\chi^2 < 10$ and the reconstructed top quark mass not to exceed 190~GeV. In measurements in the dilepton decay mode, the ATLAS and CMS Collaborations require the presence of at least one $b$-tagged jet to reduce the backgrounds. In our work, requiring one $b$ tag would not affect the reconstruction procedure, nor the kinematics, and would only decrease the sample by a factor around 0.8. As our goal is the test of the template method, we prefer to keep the full Monte Carlo statistics and therefore do not require a $b$ tag.

\subsection{Reconstruction method II}
\label{sec:recII}

This reconstruction method has been proposed and used by the ATLAS collaboration \cite{Aaboud:2019hwz} to measure top quark pair spin correlations in the dilepton channel, and it is based in the Neutrino Weighting method \cite{D0:1997pjc}. We use this reconstruction as a cross-check with reconstruction method I.

The four-momenta of the top quark and antiquark in each event can be reconstructed from the measured leptons, jets and missing transverse momentum. Since the four-momenta of the two neutrinos in the final state are not directly measured in the detector, the four-momenta of the top quark and antiquark cannot be reconstructed analytically. However, we can apply the invariant mass constraints
\begin{align}
& (p_{\ell^+} + p_{\nu})^{2}  =  m_{W}^{2} \, , \nonumber \\
& (p_{\ell^-} + p_{\bar \nu})^{2}  =  m_{W}^{2} \, , \nonumber \\
& (p_{\ell^+} + p_{\nu} + p_{b})^{2}  =  m_t^{2} \, , \nonumber \\
& (p_{\ell^-} + p_{\bar \nu} + p_{\bar b})^{2}  =  m_{t}^{2} \, .
\label{eq:inv_mass}
\end{align}
To the above four equations, we can add the following two describing the neutrino pseudorapidities:
\begin{eqnarray}
\eta_{\nu}  &=&  \arctanh \left( \frac{p_{\nu}^{z}}{\sqrt{(p_{\nu}^{x})^{2} + (p_{\nu}^{y})^{2} + (p_{\nu}^{z})^{2}}} \right) \, , \nonumber \\
\eta_{\bar \nu}  &=&  \arctanh \left( \frac{p_{\bar \nu}^{z}}{\sqrt{(p_{\bar  \nu}^{x})^{2} + (p_{\bar \nu}^{y})^{2} + (p_{\bar \nu}^{z})^{2}}} \right) \, ,
\label{eq:rap}
\end{eqnarray}
and scan over the parameters $\eta_{\nu}$ and $\eta_{\bar \nu}$ to solve the system of six equations for the neutrino four-momenta. The values of the pseudorapidities are scanned in the range $[-4.2, 4.2]$ in steps of $0.2$.

For each pair of pseudorapidities, the system of equations cannot always be solved. This can be a consequence of multiple effects, such as an incorrect choice of values for $\eta_{\nu}$ and $\eta_{\bar \nu}$, a wrong choice of the $b$ jets or detector distortions on the four-momenta of the measured particles. Furthermore, the top quarks are assumed to be produced with a mass $m_{t} = 173$ GeV, which is not always the case. In order to mitigate these effects, the top quark mass $m_{t}$ is scanned in the range $[172, 174]$ GeV in steps of $1$ GeV. Furthermore, solutions are derived for the two possible assignments of the $b$ jets. For each assignment, a Gaussian smearing is applied to the measured $p_{T}$ of each jet using a width of $13\%$ of the measured value.

For each of these combinations, the system of equations is solved to obtain two possible solutions for the four-momenta of the neutrinos. Solutions that have an imaginary component or a negative value for the energies of any of the neutrinos or top quarks are discarded. For the remaining solutions, the optimal reconstruction is defined as the one that maximises the weight:
\begin{equation}
w = \exp\left(- \frac{\Delta E_{x}^2}{2\sigma_{x}^2} - \frac{\Delta E_{y}^2}{2\sigma_{y}^2}\right) \, ,
\label{eq:E_weight}
\end{equation}
where $\Delta E_{x, y}$ is the difference between the measured missing energy in the event and the reconstructed missing energy using the four-momenta of the reconstructed neutrinos. The parameter $\sigma_{x, y}$ is a fixed scale associated to the energy resolution in the measurement of the missing energy in the event \cite{ATLAS:2018txj}. 

This reconstruction procedure is applied to events with exactly two jets. In this case, the efficiency in the reconstruction of the $t\bar{t}$ events is of $90\%$. A selection cut on the weight can be applied to increase the quality of the reconstructed events.

\section{Fit and linearity tests}
\label{sec:4}

\raggedbottom 

\begin{figure}
\begin{center}
\begin{tabular}{c}
\includegraphics[width=8.5cm,clip=]{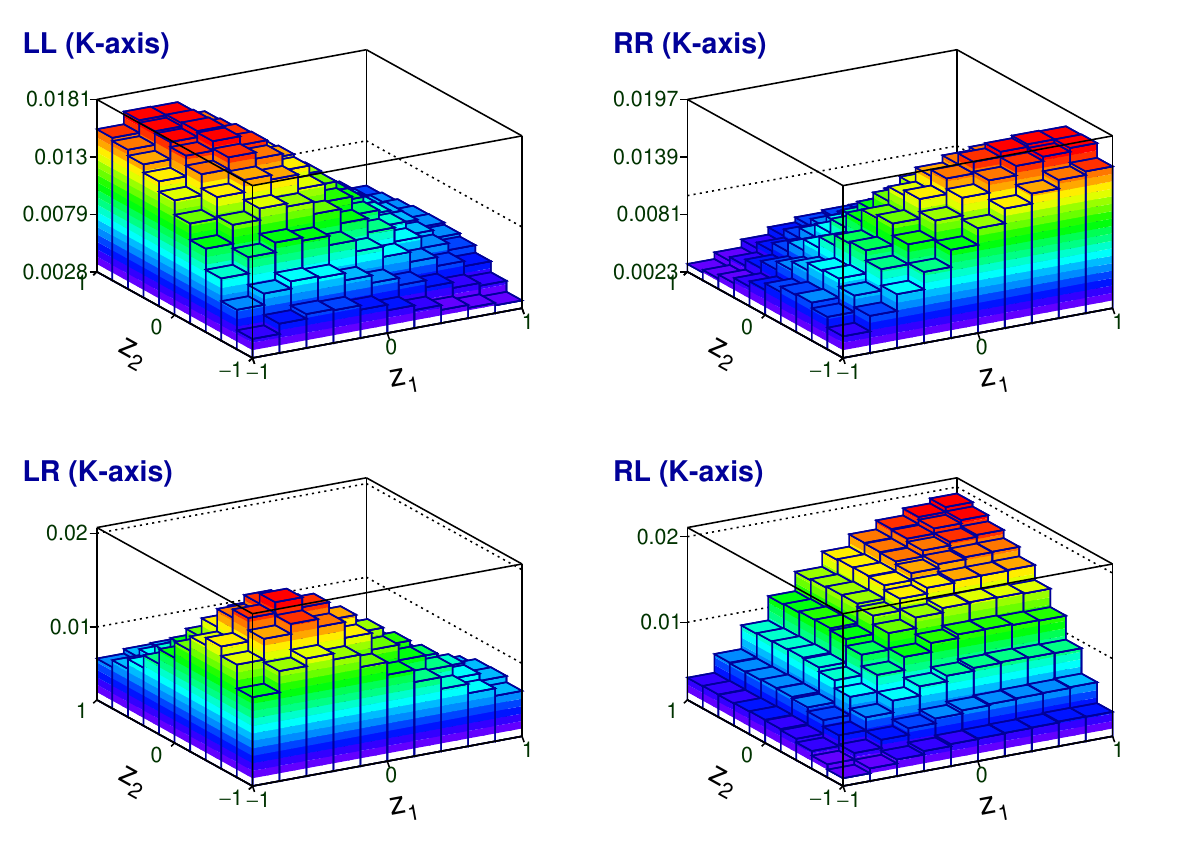} \\
\includegraphics[width=8.5cm,clip=]{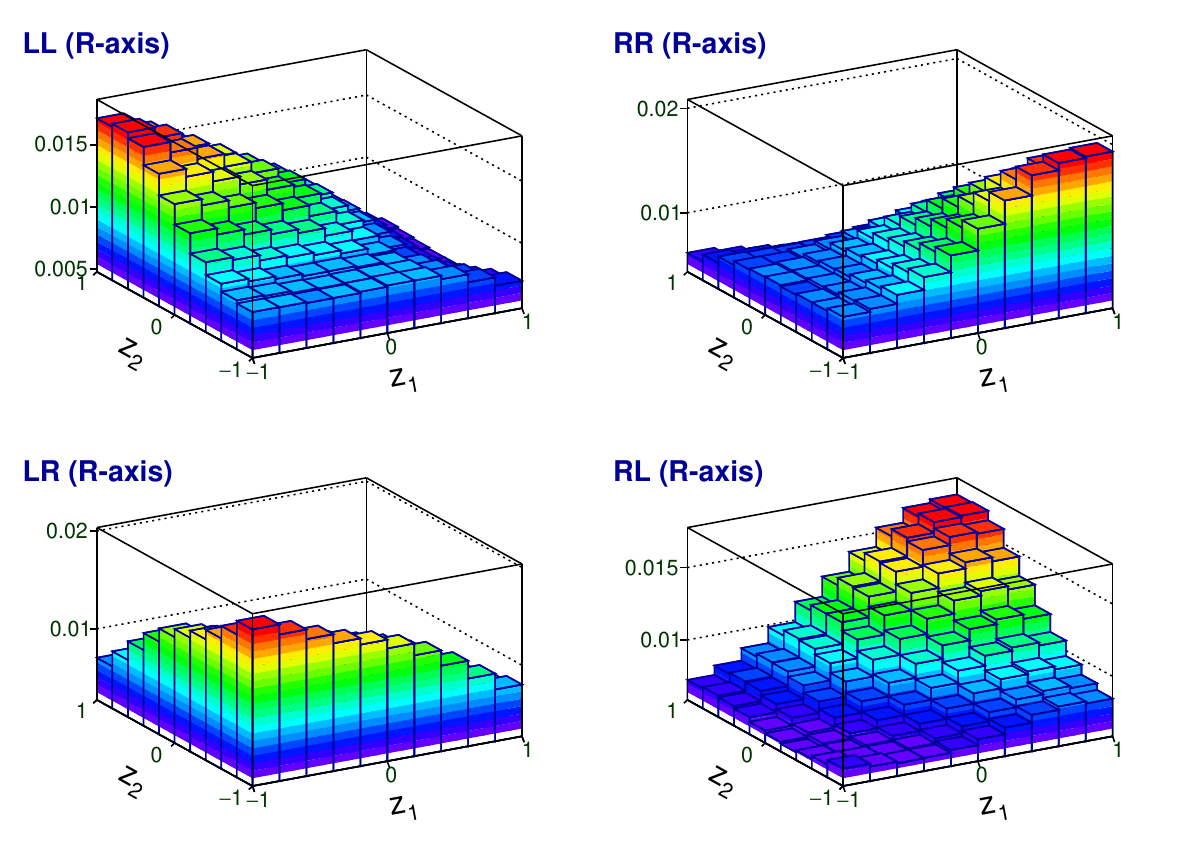} \\
\includegraphics[width=8.5cm,clip=]{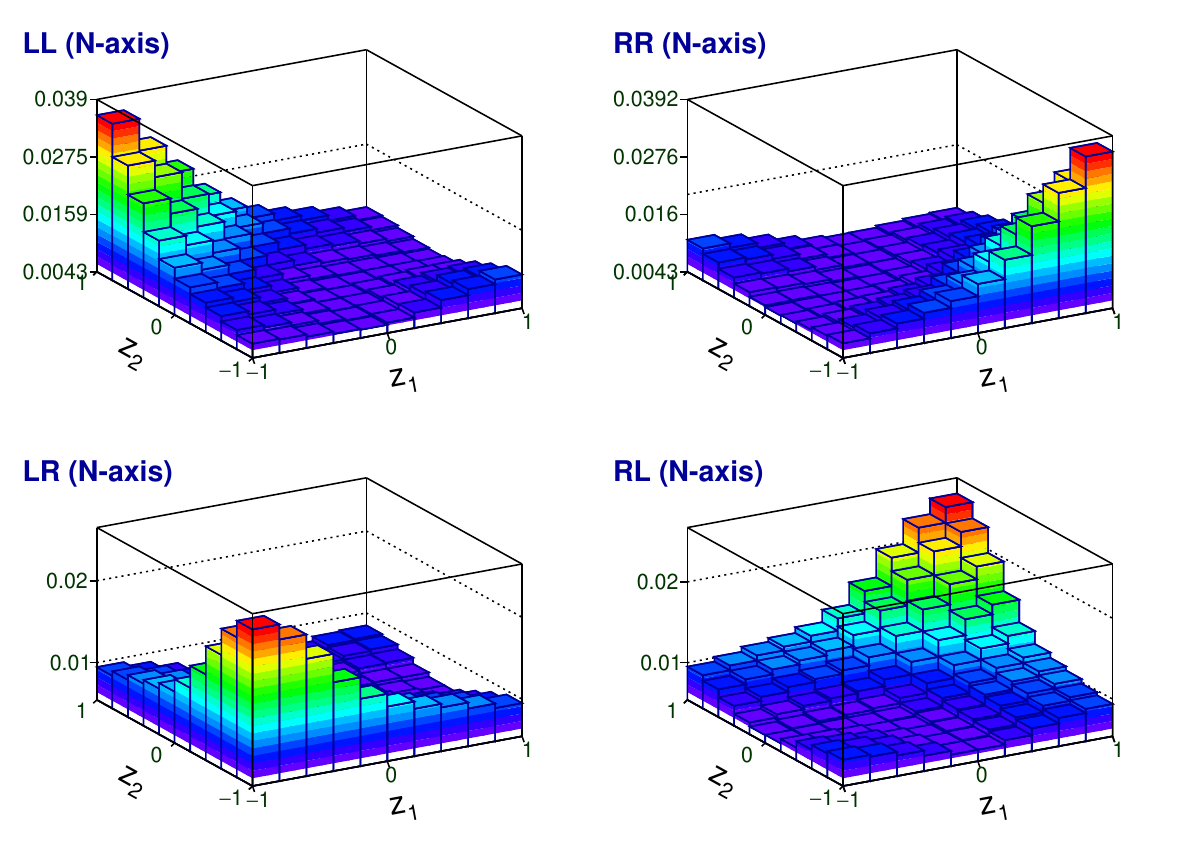}
\end{tabular}
\caption{Reconstructed normalised templates for the three sets of axes and for the $LL$, $RR$, $LR$ and $RL$ polarisations.}
\label{fig:templates}
\end{center}
\end{figure}

Both reconstruction methods give similar results, and from now on we focus on method I. The efficiencies (defined as the number of events that pass the selection and reconstruction criteria, divided by the total number of events), are shown in Table~\ref{tab:effsII}, for the samples used in this work. 
The two-dimensional templates with definite $t$ and $\bar t$ polarisation, after selection and reconstruction, are shown in Fig.~\ref{fig:templates}. These are used to obtain the optimal fit of the SM and CMDM samples, taking (\ref{ec:expansion3}) as fitting function and applying the MIGRAD optimisation algorithm of the MINUIT package~\cite{James:1975dr}, implemented in ROOT.

\begin{table}[t]
\begin{center}
\begin{tabular}{cc}
Sample & $\varepsilon$ \\[1mm]
SM & $0.174$ \\
CMDM & $0.174$
\end{tabular}

\vspace{3mm}

\begin{tabular}{cccc}
Template & \multicolumn{3}{c}{$\varepsilon$} \\
     & K-axis & R-axis & N-axis \\[1mm]
$LL$ & $0.177$ & $0.175$ & $0.178$ \\
$RR$ & $0.178$ & $0.176$ & $0.178$ \\
$LR$ & $0.160$ & $0.188$ & $0.171$ \\
$RL$ & $0.182$ & $0.161$ & $0.171$
\end{tabular}
\caption{Efficiencies for the samples and templates considered in this work. An uncertainty better than $0.001$ is evaluated, for all efficiencies.}
\label{tab:effsII}
\end{center}
\end{table}

Several tests are applied, described in what follows, to understand the performance of the template fit procedure. All tests use 2000 pseudo-experiments built from Poisson fluctuations of the bin entries in the two-dimensional distributions, allowing to build pull distributions for the correlation coefficients. For these pseudo-experiments a reference luminosity of 36.1~fb$^{-1}$ is used, to have an estimation of the statistical uncertainties in possible measurements at the LHC.

\begin{figure}
\begin{center}
\begin{tabular}{c}
\includegraphics[width=7cm,clip=]{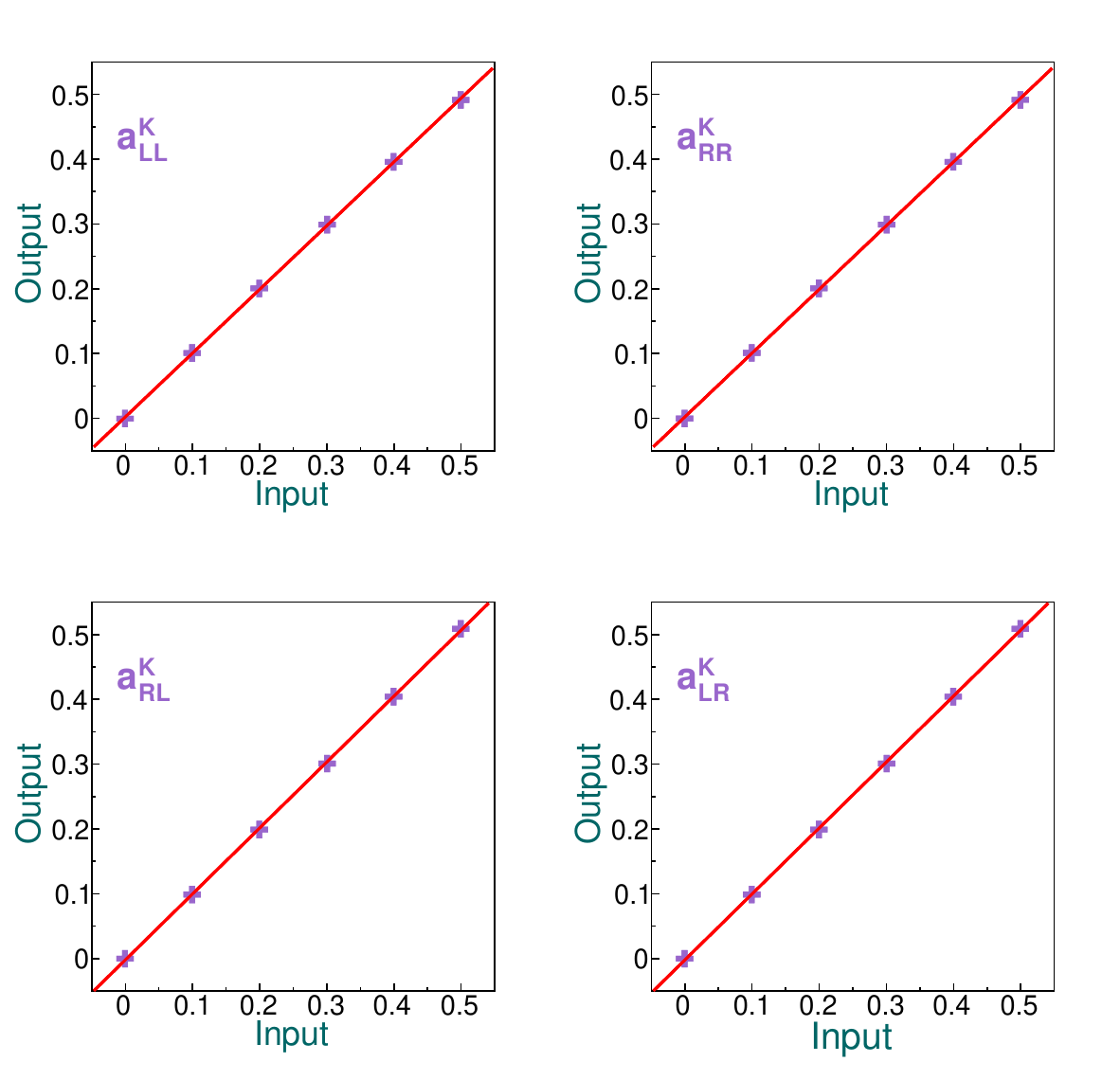} \\
\includegraphics[width=7cm,clip=]{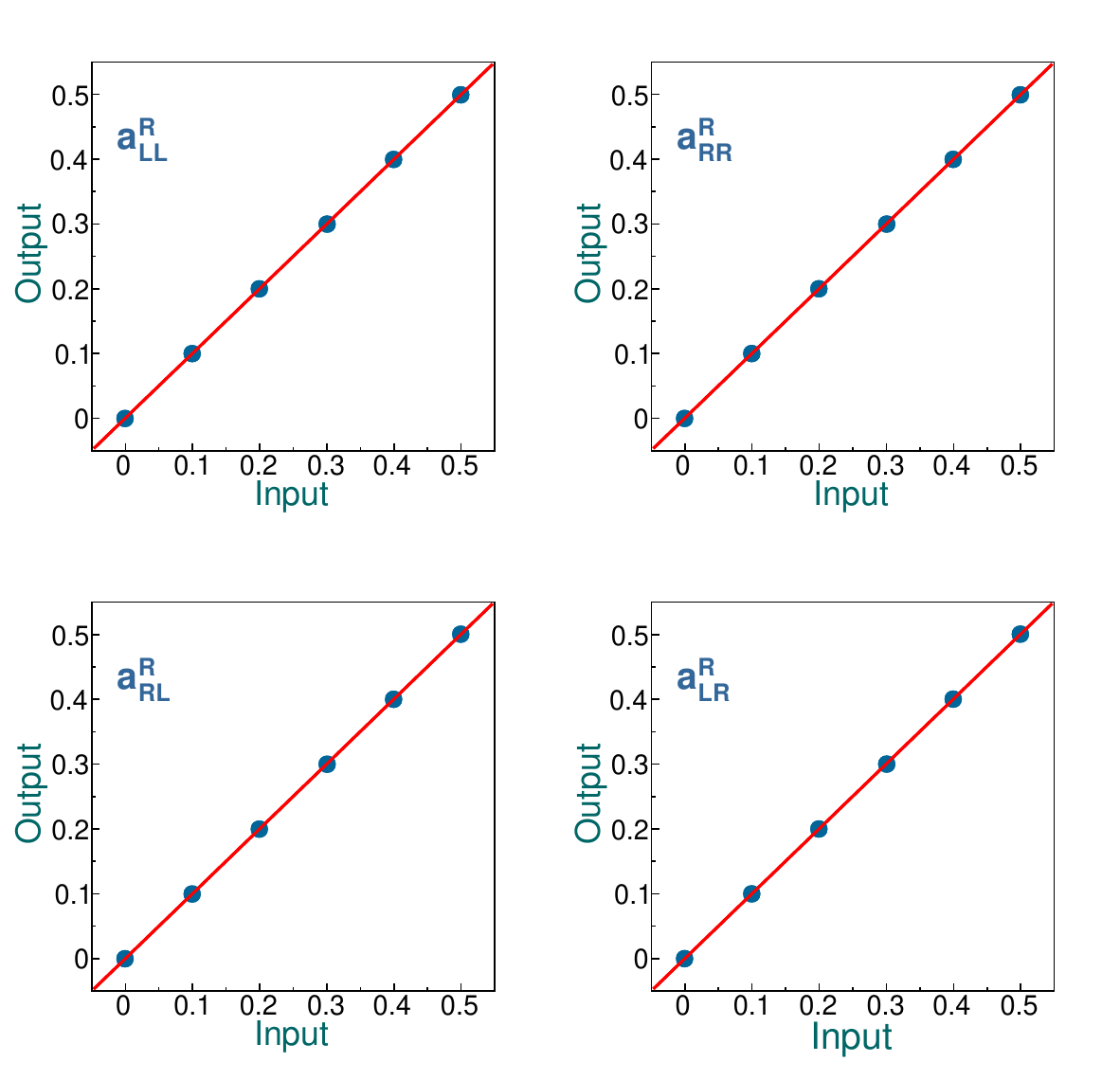} \\
\includegraphics[width=7cm,clip=]{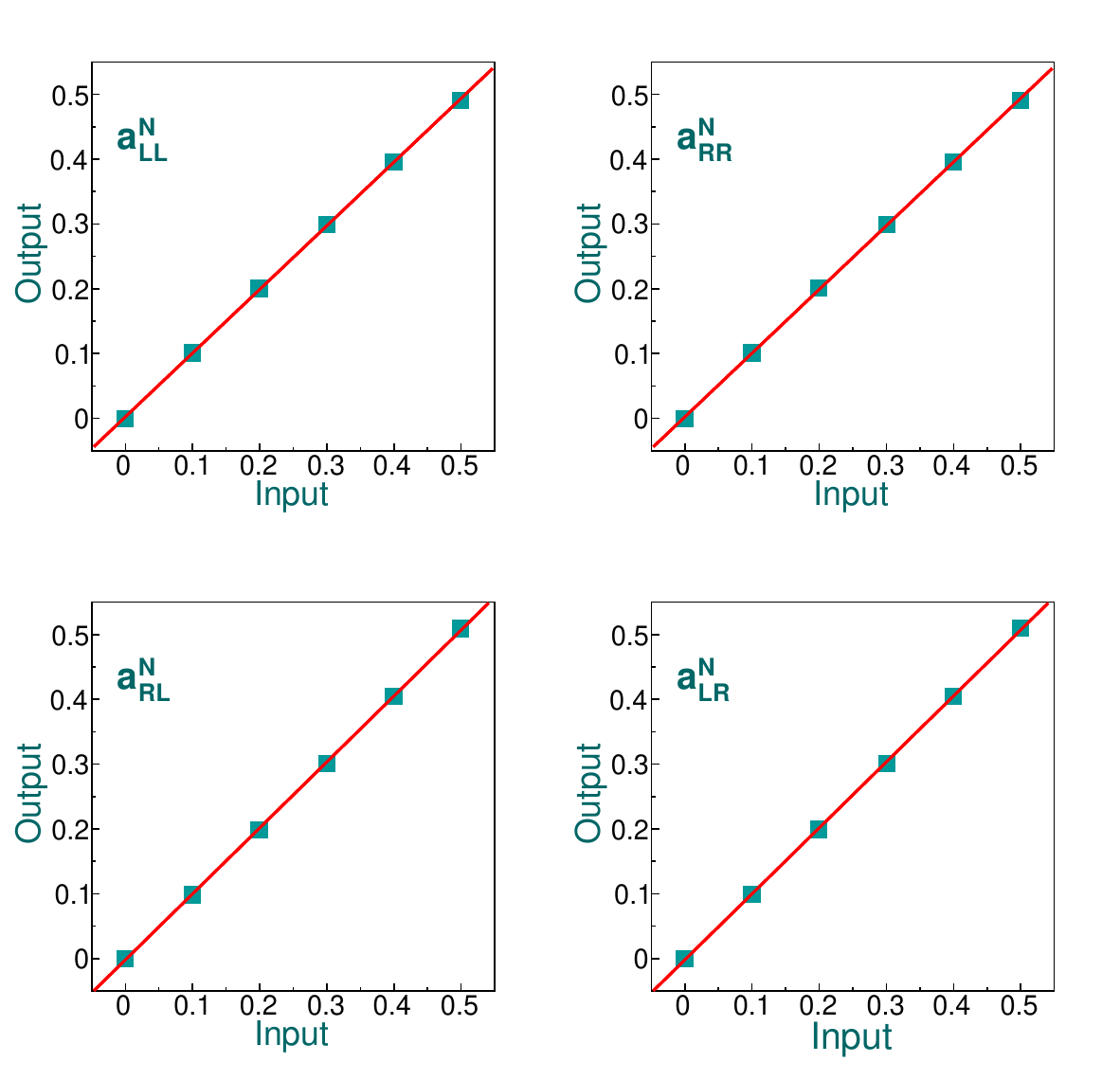}
\end{tabular}
\caption{Linearity tests for mixtures of pure $LL$, $RR$, $RL$ and $LR$ polarisations.}
\label{fig:LinearTest}
\end{center}
\end{figure}

In order to test the convergence of the fit we build, for the K, R and N sets of axes, two-dimensional distributions using mixtures of pure $LL$, $RR$, $RL$ and $LR$ templates (after event selection and reconstruction), and study the output of the fit for the correlation parameters $a_{LL}$, $a_{RR}$, $a_{RL}$ and $a_{LR}$. In this case there is no interference correction, by construction. For simplicity, we choose $a_{LL}=a_{RR}$ and $a_{RL}=a_{LR}$ ranging from 0 to 0.5, with a 0.1 step size. In Figure~\ref{fig:LinearTest} the linearity tests are shown for the different correlations parameters and the three sets of axes. Good correlations are observed between input and fitted values.

\section{Extracting $t \bar t$ polarisations}
\label{sec:5}

\begin{figure}[t]
\begin{center}
\begin{tabular}{c}
\includegraphics[width=6cm,clip=]{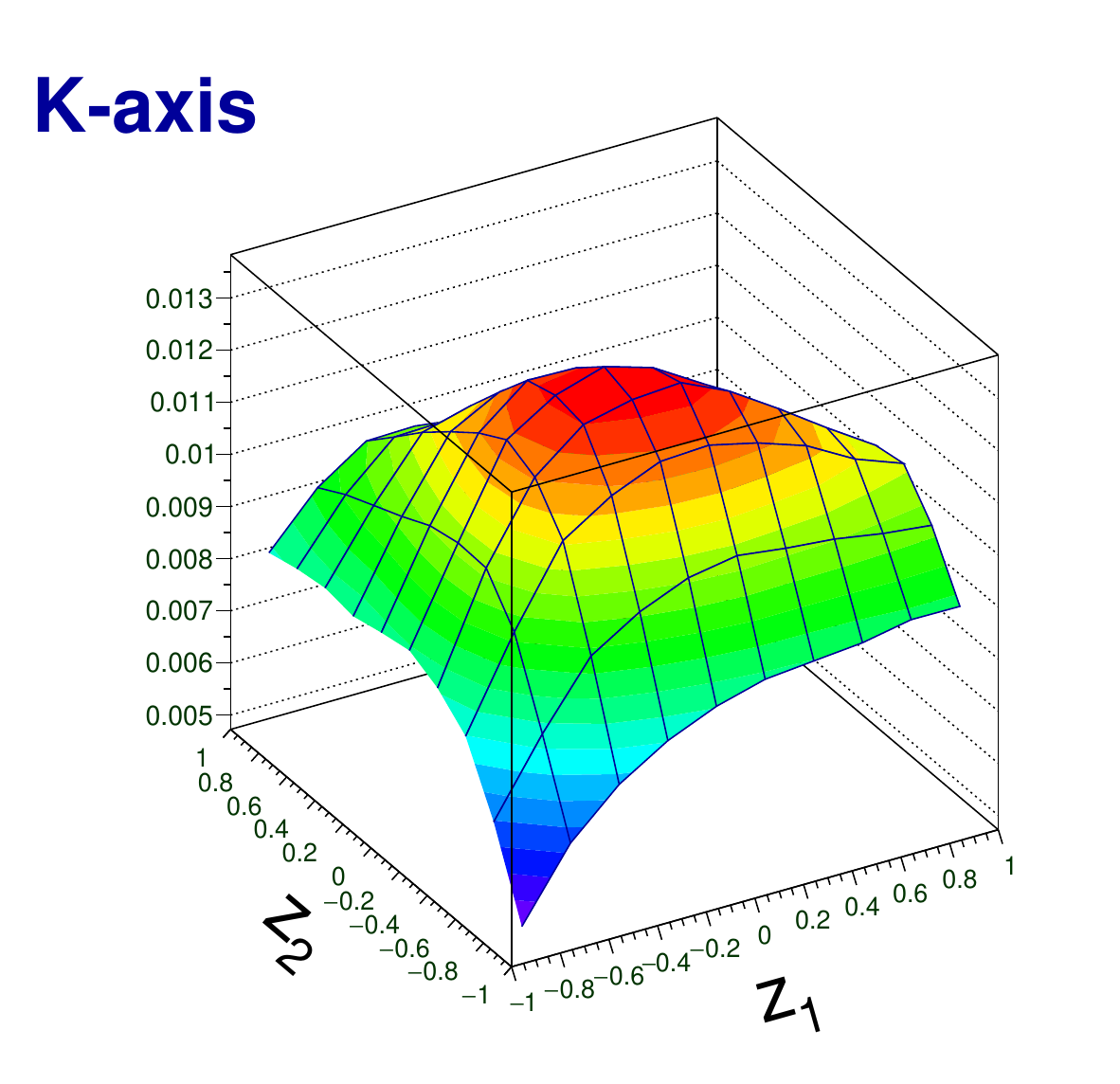} \\
\includegraphics[width=6cm,clip=]{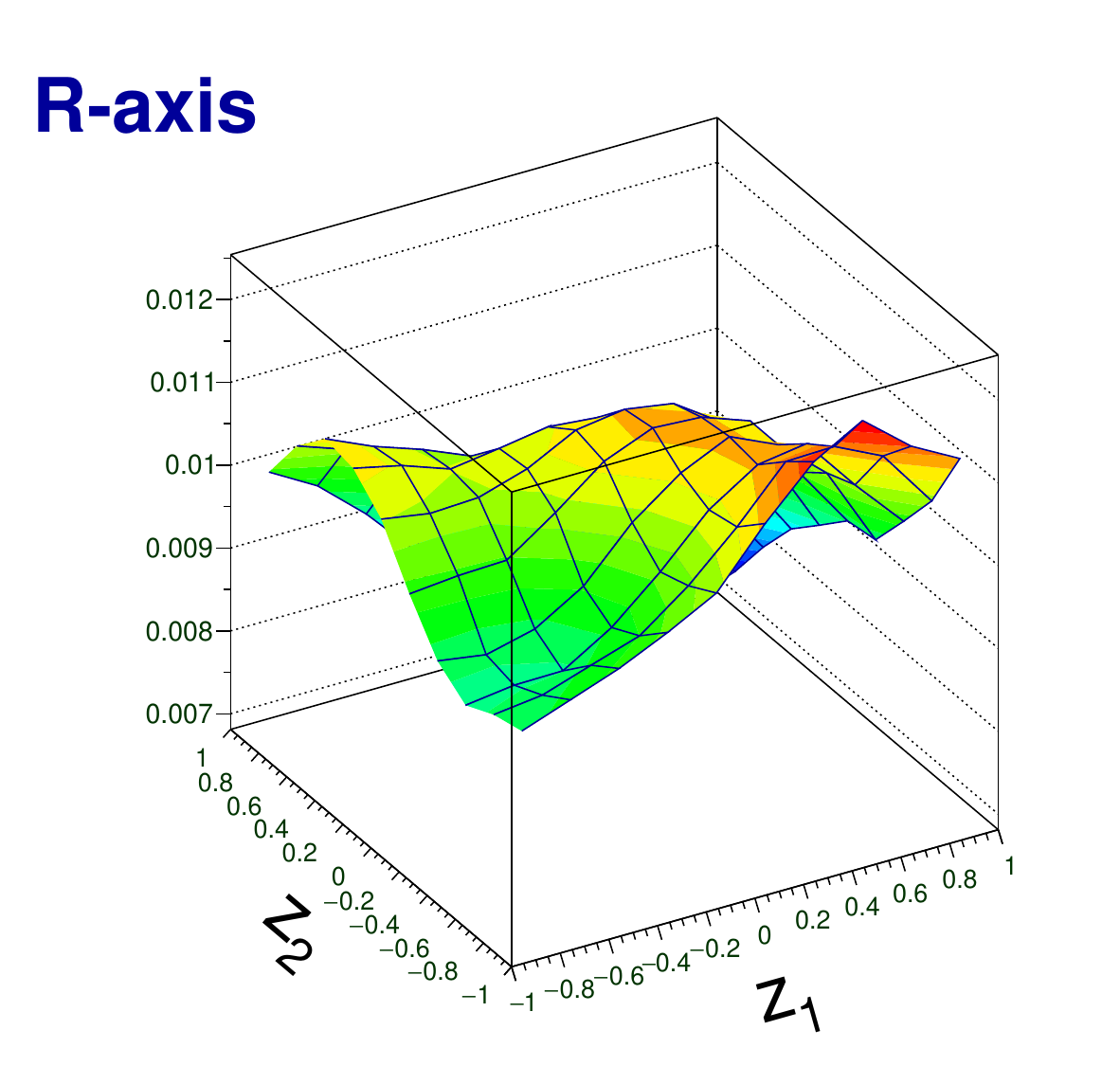} \\
\includegraphics[width=6cm,clip=]{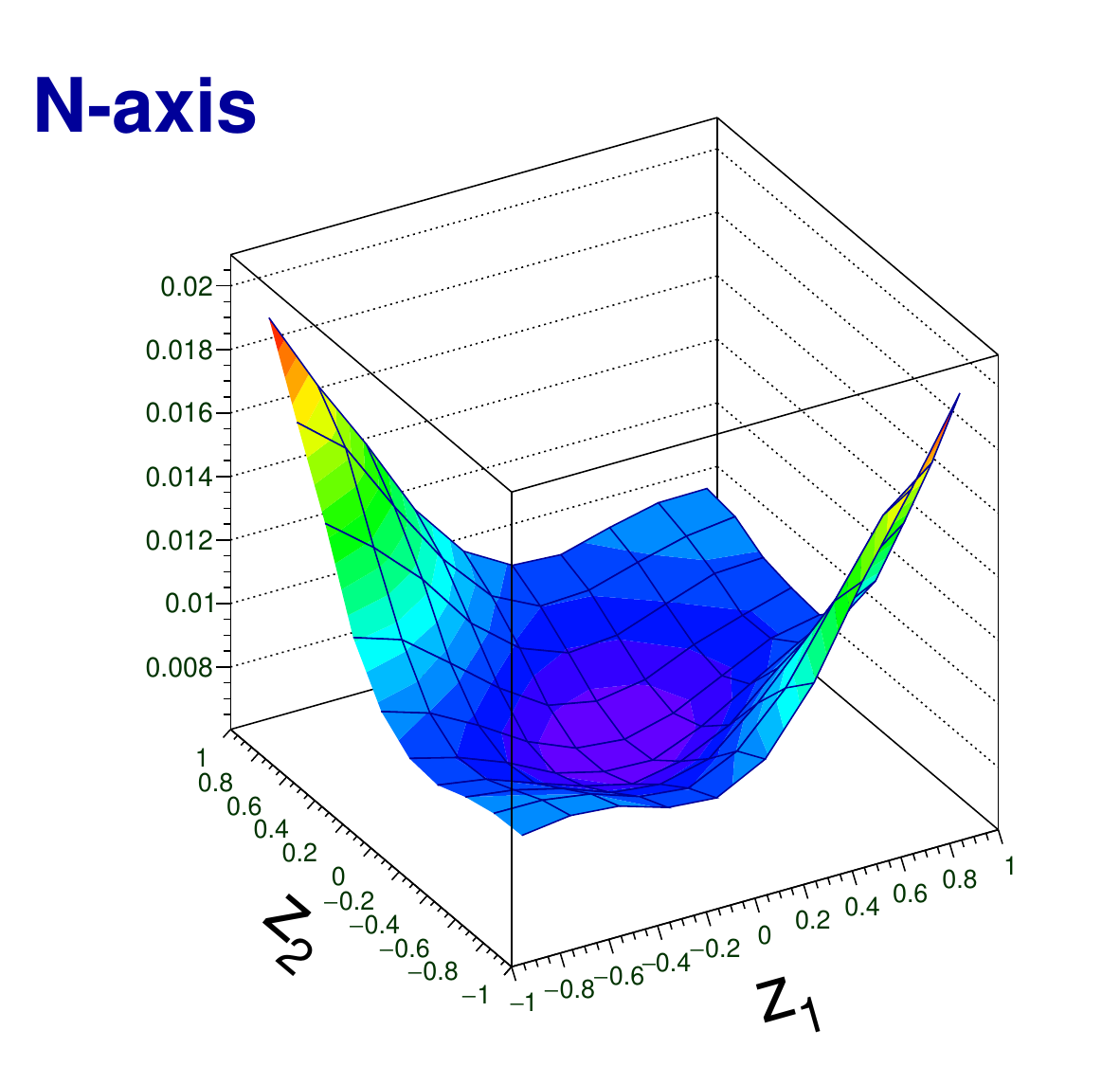}
\end{tabular}
\caption{Normalised two-dimensional distributions for the SM sample using the K (top), R (middle) and N (bottom) sets of axes.}
\label{fig:SMfitsII}
\end{center}
\end{figure}

\begin{figure}
\begin{center}
\begin{tabular}{c}
\includegraphics[width=6cm,clip=]{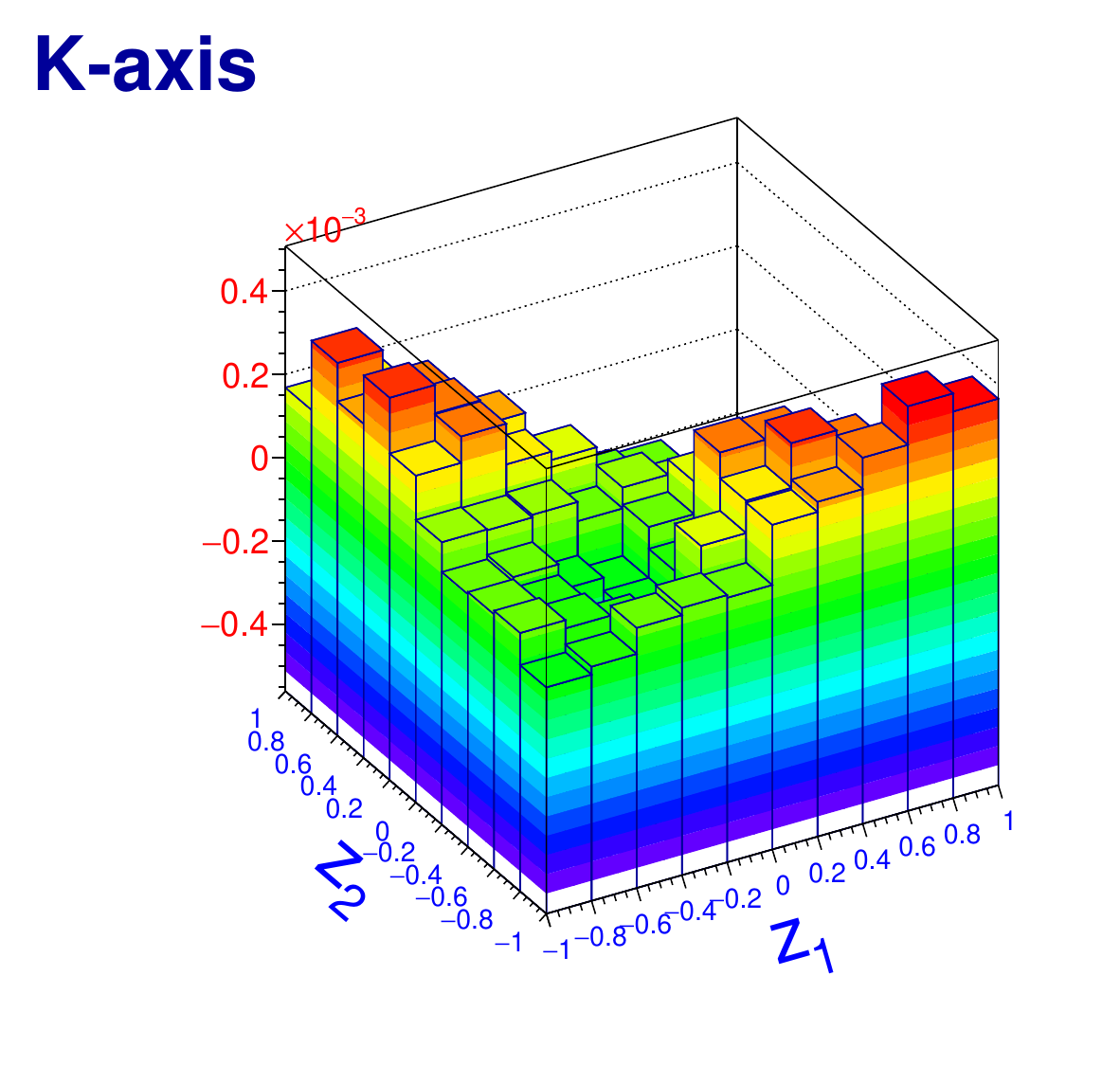} \\
\includegraphics[width=6cm,clip=]{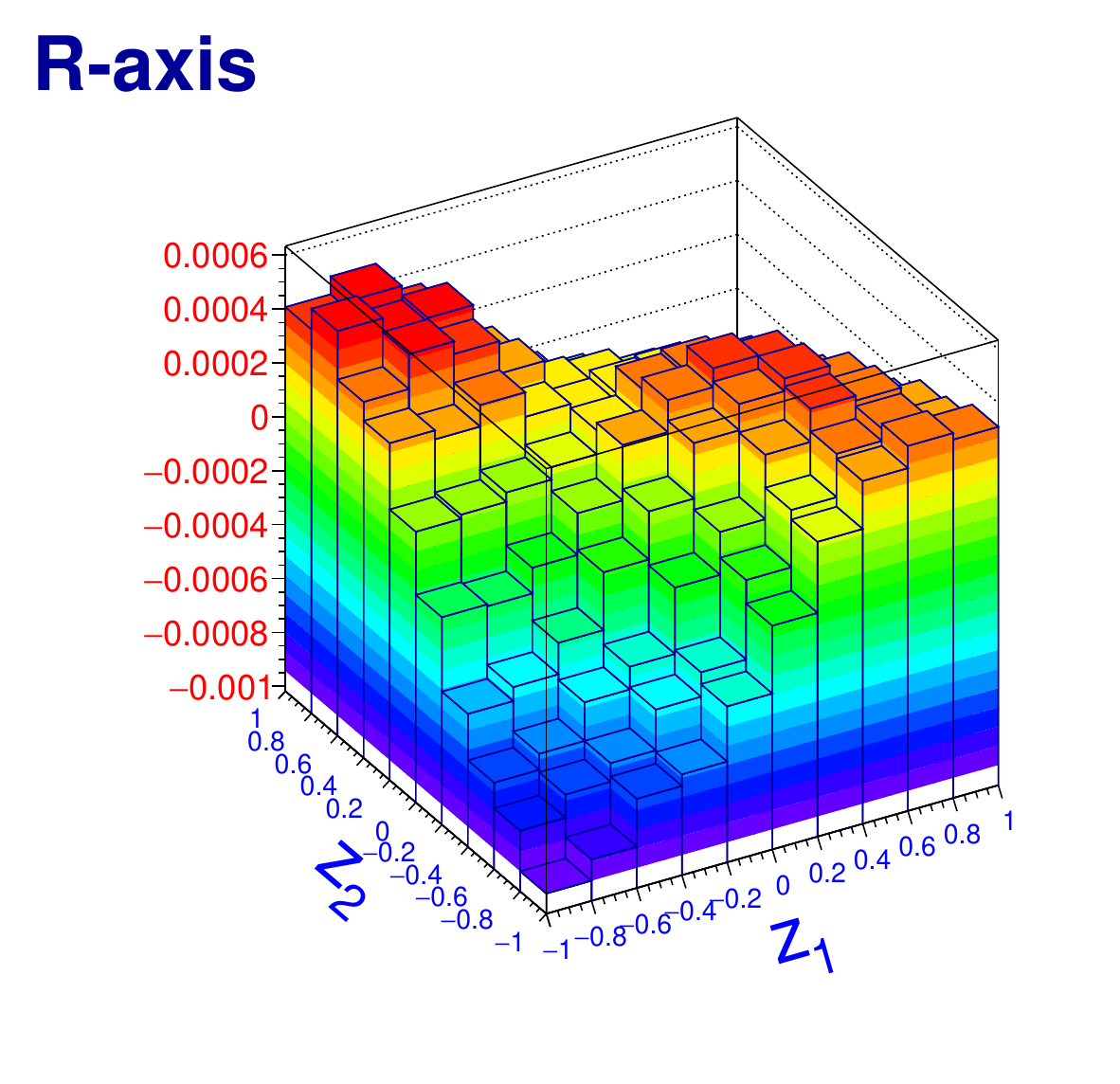} \\
\includegraphics[width=6cm,clip=]{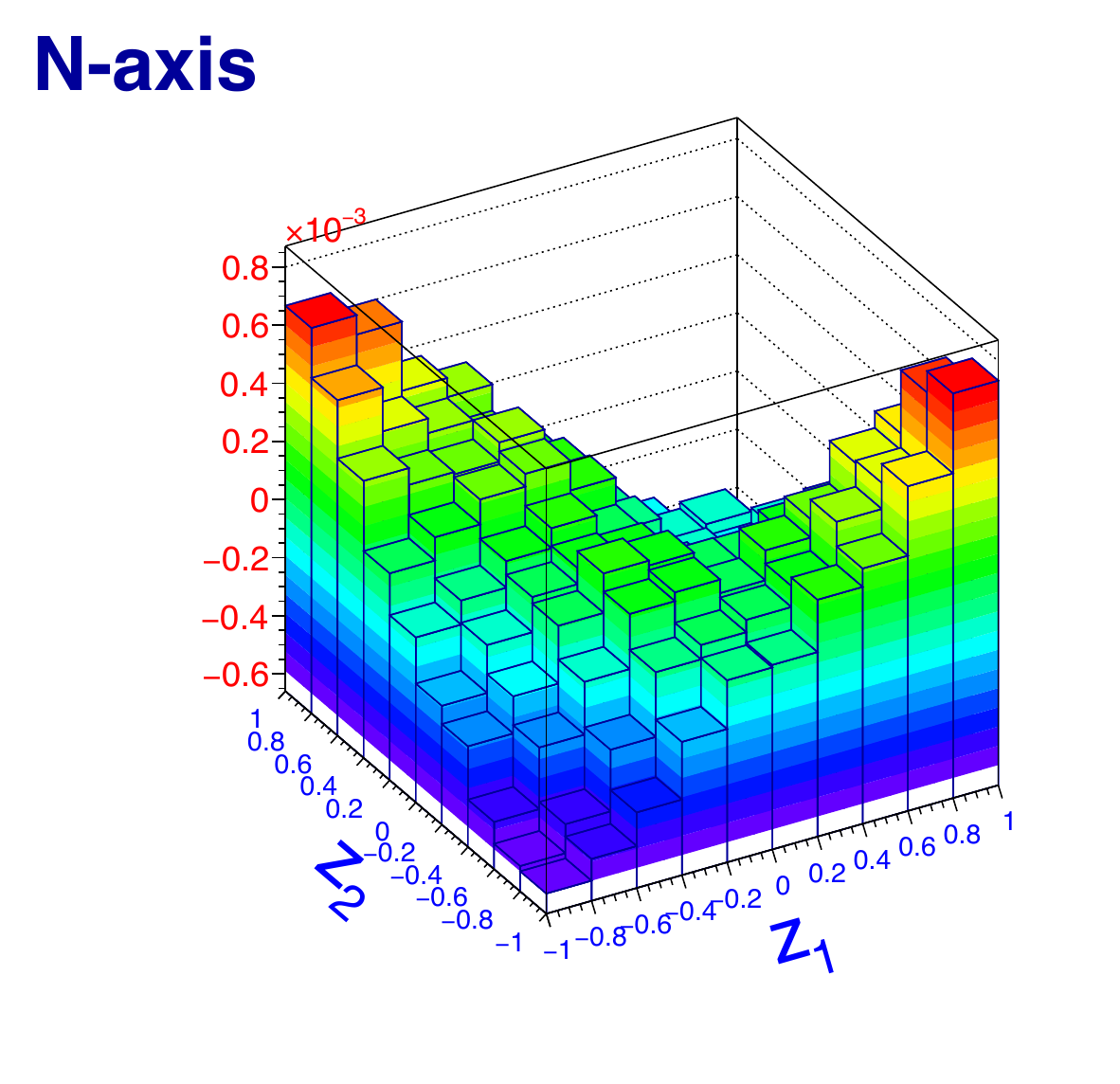} \\
\end{tabular}
\caption{Interference corrections $\Delta_\text{int}$ for the $K$ (top), $R$ (middle) and $N$ (bottom) axes, normalised to the total number of events in the SM sample.}
\label{fig:migrationsII}
\end{center}
\end{figure}

Here we investigate the extraction of the polarisation parameters in the SM as well as in the CMDM samples. For illustration, we show in Fig.~\ref{fig:SMfitsII} the normalised two-dimensional distributions for the SM sample, using the three sets of axes.
As discussed in section~\ref{sec:2}, a residual interference correction $\Delta_\text{int}$ may arise from imperfect reconstruction of the lepton angles, which makes the interference between polarisation states non-negligible. When comparing theory with data, it has to be taken into account. By subtracting the SM true two-dimensional distribution (after simulation and reconstruction) from the distribution obtained with a SM-like mixture of $LL$, $RR$, $LR$ and $RL$ distributions, weighted with corresponding polarisation coefficients, we can determine the corrections, which are shown in Fig.~\ref{fig:migrationsII}. For the computation of this correction one third of the SM sample is used, corresponding to around $10^5$ events. (These events are not used in the subsequent fit.) The inclusion of this term allows to perform the desired template fit, using (\ref{ec:expansion3}) to extract the polarisation coefficients $a_{XX'}$, not only in the SM sample but also in presence of a large CMDM.
The values of the polarisation coefficients are obtained by performing 2000 pseudo-experiments, in which the entries in the binned two-dimensional distributions are allowed to fluctuate following a Poisson distribution, assuming a luminosity of 36.1 fb$^{-1}$. In each pseudo-experiment Eq.~(\ref{ec:expansion3}) is used to obtain the coefficients with a fit. Results are summarised in Table~\ref{tab:coeffs}, together with the predictions directly calculated from the SM and CMDM Monte Carlo samples, at parton level. The spin correlation coefficients
\begin{equation}
C = a_{LL} + a_{RR} - a_{LR} - a_{RL}
\end{equation}
and $t$, $\bar t$ polarisations
\begin{eqnarray}
P_t & = & a_{RR} + a_{RL} - a_{LR} - a_{LL} \,, \notag \\
P_{\bar t} & = & a_{RR} + a_{LR} - a_{RL} - a_{LL} \,,
\end{eqnarray}
are also included for completeness, for the three sets of axes. The spin correlation coefficients and polarisations are calculated from the polarisation coefficients $a_{XX'}$ in each pseudo-experiment.

\flushbottom
 
The fit results presented in Table~\ref{tab:coeffs} correspond to the mean and standard deviation of the distribution obtained with the 2000 pseudo-experiments. This `fit' statistical uncertainty represents the statistical uncertainty that would be expected in a measurement with $L = 36.1~\text{fb}^{-1}$. It is independent of the Monte Carlo statistical uncertainty in the samples, which have around $2 \times 10^5$ events.
\begin{table}[t]
\begin{center}
\begin{scriptsize}
\begin{tabular}{ccccc}
K & \multicolumn{2}{c}{SM} & \multicolumn{2}{c}{CMDM} \\
 & Prediction & Fit  & Prediction & Fit \\
$a_{LL}$ & $0.335 \pm 0.001$ & $0.337 \pm 0.006$          & $0.349 \pm 0.001$ & $0.350 \pm 0.006$\\

$a_{RR}$ & $0.336 \pm 0.003$ & $0.330 \pm 0.005$          & $0.349 \pm 0.001$ & $0.339 \pm 0.005$\\

$a_{LR}$ & $0.165 \pm 0.003$ & $0.167 \pm 0.007$          & $0.151 \pm 0.001$ & $0.175 \pm 0.007$\\

$a_{RL}$ & $0.165 \pm 0.002$ & $0.160 \pm 0.004$          & $0.151 \pm 0.001$ & $0.131 \pm 0.004$\\

$C_{\text{kk}}$ & $0.340 \pm 0.002$ & $0.340 \pm 0.019$   & $0.394 \pm 0.004$ & $0.383 \pm 0.019$\\

$P_{t}$         & $0.001 \pm 0.002$ & $-0.014 \pm 0.008$  & $0.000 \pm 0.001$ & $-0.058 \pm 0.008$\\

$P_{\bar{t}}$   & $0.001 \pm 0.002$ & $0.000 \pm 0.008$   & $0.001 \pm 0.002$ & $0.033 \pm 0.008$

\end{tabular}

\vspace{5mm}

\begin{tabular}{ccccc}
R & \multicolumn{2}{c}{SM} & \multicolumn{2}{c}{CMDM} \\
 & Prediction & Fit  & Prediction & Fit \\
$a_{LL}$ & $0.258 \pm 0.001$ & $0.254 \pm 0.006$          & $0.290 \pm 0.002$ & $0.291 \pm 0.006$\\

$a_{RR}$ & $0.259 \pm 0.002$ & $0.264 \pm 0.006$          & $0.289 \pm 0.002$ & $0.290 \pm 0.006$\\

$a_{LR}$ & $0.242 \pm 0.001$ & $0.236 \pm 0.006$          & $0.210 \pm 0.001$ & $0.210 \pm 0.006$\\

$a_{RL}$ & $0.241 \pm 0.002$ & $0.241 \pm 0.006$          & $0.211 \pm 0.001$ & $0.201 \pm 0.006$\\

$C_{\text{rr}}$ & $0.036 \pm 0.002$ & $0.041 \pm 0.019$   & $0.159 \pm 0.002$ & $0.170 \pm 0.019$\\

$P_{t}$         & $0.0004 \pm 0.0005$ & $0.015 \pm 0.010$   & $-0.001 \pm 0.004$ & $-0.011 \pm 0.010$\\

$P_{\bar{t}}$   & $0.002 \pm 0.002$ & $0.006 \pm 0.010$   & $-0.001 \pm 0.003$ & $0.008 \pm 0.009$\\
\end{tabular}
\vspace{5mm}

\begin{tabular}{ccccc}
N & \multicolumn{2}{c}{SM} & \multicolumn{2}{c}{CMDM} \\
 & Prediction & Fit  & Prediction & Fit \\
$a_{LL}$ & $0.333 \pm 0.001$ & $0.329 \pm 0.004$          & $0.358 \pm 0.001$ & $0.363 \pm 0.004$\\

$a_{RR}$ & $0.334 \pm 0.002$ & $0.329 \pm 0.004$          & $0.358 \pm 0.001$ & $0.352 \pm 0.004$\\

$a_{LR}$ & $0.166 \pm 0.001$ & $0.164 \pm 0.004$          & $0.142 \pm 0.0003$ & $0.138 \pm 0.004$\\

$a_{RL}$ & $0.167 \pm 0.002$ & $0.169 \pm 0.004$          & $0.142 \pm 0.001$ & $0.136 \pm 0.004$\\

$C_{\text{nn}}$ & $0.336 \pm 0.002$ & $0.325 \pm 0.010$   & $0.433 \pm 0.002$ & $0.442 \pm 0.010$\\

$P_{t}$         & $0.002 \pm 0.001$ & $0.005 \pm 0.009$   & $-0.001 \pm 0.002$ & $-0.014 \pm 0.009$\\

$P_{\bar{t}}$   & $0.000 \pm 0.002$ & $-0.005 \pm 0.008$   & $0.000 \pm 0.001$ & $-0.009 \pm 0.009$\\
\end{tabular}
\end{scriptsize}
\caption{Theory predictions and best-fit values for various polarisation coefficients in the SM and CMDM data samples. For the theory predictions, the quoted uncertainty corresponds to the Monte Carlo statistical uncertainty.}
\label{tab:coeffs}
\end{center}
\end{table}

Both in the SM and CMDM samples the quantities extracted from the fit are close to the parton-level predictions, extracted directly from the LO event sets. Remarkably, this is also the case for the CMDM sample for which the $\Delta_\text{int}$ correction is evaluated within the SM. (We have also checked the evaluation with the CMDM sample itself and the results are compatible.) The small shifts observed in $a_{LR}$ and $a_{RL}$ for the K axis seem to have a statistical origin. We remark that this benchmark has large deviations in the spin correlation coefficients, which make it experimentally excluded~\cite{Aguilar-Saavedra:2018ggp}. The point of having this benchmark point is precisely to test the robustness of the method for large deviations. Therefore, the evaluation of $\Delta_\text{int}$ in the SM seems to have a quite wide range of validity.
Overall, the results in Table~\ref{tab:coeffs} show that already with existing data the measurements of $a_{XX'}$ are promising.

\section{Discussion}
\label{sec:6}

The template method for the measurement of the polarisation fractions $a_{XX'}$ offers an alternative to existing methods~\cite{CMS:2019nrx,Aaboud:2019hwz} which, in addition of the $t$ and $\bar t$ net polarisation and spin correlation coefficients, allows for individual measurements of the $a_{XX'}$ coefficients. Using a fast simulation we have shown that the measurements are feasible, and the fit to the templates accurately recovers the parton-level values for $a_{XX'}$. Our results demonstrate that the template method could already be used by the ATLAS and CMS Collaborations with existing data. 

We have not addressed systematic uncertainties which necessarily have to be studied within the context of an experimental analysis. We have estimated the statistical uncertainty in the measurements, using pseudo-experiments in which the different distributions are subject to random fluctuations using Poisson distributions. Systematic uncertainties can easily be incorporated into the pseudo-experiments, by changing parameters such as energy scales.

The template method seems quite robust. The imperfect reconstruction of $t$ and $\bar t$ momenta, which generates residual interference corrections, can easily be taken into account by generating a SM sample and comparing with a SM-like combination of templates without interference. (In the limit of perfect reconstruction, this correction vanishes as shown in Section~\ref{sec:2}.) This small correction $\Delta_\text{int}$ can be calculated within the SM, and we have found that it can also be used to extract the  $a_{XX'}$ coefficients in a sample with a large top CMDM, in which the deviations in the spin correlations are sizable, so as to have this benchmark experimentally excluded.

The template method is based on the generation of $t \bar t$ samples with definite polarisation. Higher-order corrections affect very weakly the lepton angular distributions, which are the ones that we use to extract the $a_{XX'}$ coefficients. They also modify the $t \bar t$ kinematics, and hence the acceptance. Although a generation of templates at NLO is desirable, a kinematical reweighting of LO samples may be sufficient. We encourage the ATLAS and CMS Collaborations to further investigate the feasibility of the measurements using the template method.

\section*{Acknowledgements}

 This work is supported by the Spanish Research Agency (Agencia Estatal de Investigacion) through the grant IFT Centro de Excelencia Severo Ochoa SEV-2016-0597, by the projects PID2019-110058GB-C21, PID2019-110058GB-C22 from MICINN/AEI/ERDF,
 and by projects CERN/FIS-PAR/0004/2019, CERN/FIS-PAR/ 0029/2019 from FCT.
 The work of P.M.R. is supported through the FPI grant BES-2016-076775. The work of M.C.N.F. was supported by the PSC-CUNY Awards 63096-00 51 and 64031-00 52.

\bibliographystyle{utphys}
\bibliography{references}

\end{document}